\documentclass[aps,prl,preprint]{revtex4}

\pdfoutput=1
\usepackage{graphicx}
\usepackage{subfigure}
\usepackage{verbatim}
\usepackage{color}
\usepackage{amsmath}

\begin{document}

\title{Hierarchical Self-Assembly of Asymmetric Amphiphatic Colloidal Particles}
\author{William L. Miller}
\author{Angelo Cacciuto}
\email{ac2822@columbia.edu}
\affiliation{Department of Chemistry, Columbia University, \\New York, New York 10027}
\date{\today}

\begin{abstract}
From dumbbells to FCC crystals, we  study the self-assembly pathway 
of amphiphatic, spherical colloidal particles as a function of the size of the hydrophobic region using  molecular dynamics simulations.  Specifically, we analyze
how local inter-particle interactions correlate to the final self-assembled aggregate and how they affect the dynamical pathway of structure formation.
We present a detailed diagram separating the many phases that we find for different sizes of the hydrophobic area, and uncover a narrow region where particles self-assemble into hollow, faceted
cages that could potentially find interesting engineering applications.

\end{abstract}

\maketitle

\section{Introduction}
Spontaneous assembly of components into large ordered aggregates is a ubiquitous phenomenon in nature, and is observed across all length scales. Aggregation of proteins into functional nanomachines~\cite{Alberts}, formation of viral capsids~\cite{Watson,Klug}, packing of phospholipids into biological membranes~\cite{Sackmann} and assembly of colloidal particles into macroscopic  crystals~\cite{Subramanian} are just a few manifestations of this fundamental process. Because of advances in particle synthesis~\cite{DeVries,Schnablegger,Hong,Weller,Hobbie}, it is now possible to produce colloidal particles that are anisotropic both in shape and surface chemistry, thus providing an unlimited variety of building blocks that can spontaneously assemble into an unprecedented number of structures holding promise for the development of materials with novel functional, mechanical, and optical properties.

Although the generic features of particle aggregation can be described, at least phenomenologically, in terms of simple thermodynamic arguments~\cite{Israelachvili,Zhang,Leckband,Nagarajan}, the details of the process are far from being understood. 
In fact, for self-assembly to take place, a very delicate balance between entropic and energetic contributions, coupled to a precise geometric character of the components, must be satisfied. In general, self-assembly is not to be expected unless a careful design of the building blocks has been performed beforehand,
and this has inspired a large body of work dedicated to gain insight into how the geometry of the interparticle interactions and the shape of the particles themselves  determine the dynamical pathway and structure they aggregate into.  The ultimate goal is to be able to gather a sufficient understanding of the forward self-assembly process to then be able to develop tools that will allow us to tailor interparticle interactions to target desired structures. With this in mind, there is a clear necessity  to explore a new dimension in the classical temperature/concentration phase diagrams: the geometry of the interactions (see reference~\cite{glotzer2} for a recent perspective on the subject and a review of the relevant literature.) 

Recently, the self-assembly pathway of Janus particles (spherical particles with one hemisphere hydrophobic and the other hydrophilic) has been described in some detail~\cite{Hong2} both experimentally and numerically.  Janus particles are important because they represent what is probably the simplest model where 
one can explicitly study and ask fundamental questions about the role of surface anisotropy in colloidal aggregation. What was found is a rich behavior in terms of both the dynamics of self-assembly, and the final structure of the aggregates. Specifically,
by tuning the repulsion between the hydrophilic hemispheres, one can drive the system into two distinct phases. In one phase  particles form a gas of small  clusters composed of 4 to 13 particles, in the other phase, at large salt concentration, particles organize into long, branched, worm-like structures formed by cooperative fusion of the small clusters.  {Such particles are a particular type of a wider class of patchy particles, the self-assembly of a wide variety of which has been studied previously~\cite{glotzer,sciortino}.}

Inspired by these experimental results, in this paper we go one step forward and explore dynamics and structure formation of anisotropic 
amphiphatic particles. While the dividing surface  between hydrophobic and hydrophilic regions in
Janus particles is set at the equator, we shall consider particles in which this boundary is located at an arbitrary
latitude on the particle's surface, and we systematically study how, at a fixed concentration, these particles self-assemble depending on the
size of the hydrophobic region.  {We thus explore a key ``anisotropy dimension'' proposed in \cite{glotzer2} as related to the self-assembly of these particles.}

This system is quite interesting because one can easily predict  
the formation of a wealth of different structures  as a function of the location of the dividing surface. 
For instance, we know that when the hydrophobic  region covers a small area, particles  can only assemble into dumbbells. 
We know that when the dividing surface is placed to slightly larger values, so that more particles 
can share the same hydrophobic surface, particles condense into small stable  clusters. We also know the structures resulting from self-assembly of Janus particles, 
and of course, in the limit  of full coverage, we recover an isotropic potential 
which is known to lead to the formation of an FCC crystal. Using this simple system we can study 
systematically the formation of aggregates whose structure ranges from zero dimensions
(dumbbells and meso-particles) to three dimensions (FCC crystals), and we can analyze 
how  the specificity of the local inter-particle interactions correlates to the final self-assembled structure and its dynamics.

\section{Methods}
We model the amphiphatic  character of the particles  via an interaction potential
that depends on both the separation between particle surfaces and the angle between particle axes, so that a precise shape and extent of the interaction may be defined.  
Our choice of the interaction potential is inspired by the model introduced in reference~\cite{Hong2} that has been used to analyze actual experimental data of Janus Particles.

{
The potential presented in~\cite{Hong2} was specifically tailored to describe the physical properties of Janus particles at different salt concentrations, $\rho_s$. At low and intermediate salt concentrations, 
the repulsion between the charged hemispheres constrains  particles 
to interact head-on, with angular deviations that depend in a nontrivial way on $\rho_s$. However,
the more interesting structure formation (worm-like clusters) appears at large salt concentrations, where the role of electrostatic interactions becomes negligible. In this limit, particles can freely rotate, once in contact, within the boundaries of the hydrophobic regions.  These regions can therefore be appropriately described by a short-ranged, 
isotropic attractive potential that acts within the boundaries of that region. Similarly, the hydrophilic region is well-characterized by a simple short-ranged repulsive interaction. 
The potential used in our paper reflects this phenomenological behavior and has the following form:}

\[V(r,\theta_1,\theta_2) = V_{\rm rep}(r) + V_{\rm att}(r)\phi(\theta_1)\phi(\theta_2)\,,\]
where $r$ is the distance between particles, $\theta_1$ is the angle between the axis of particle 1 
and the axis between particle centers, and $\theta_2$ is the analogous angle for particle 2.  
$V_{\rm rep}(r)$ is a symmetric repulsive interaction that accounts for the particle excluded volume, and
has the form of a shifted-truncated  Lennard-Jones potential:
\[V_{\rm rep}(r) = \begin{cases} 4\epsilon_0 \left[ (\frac{\sigma}{r})^{12} - (\frac{\sigma}{r})^{6} +\frac{1}{4}\right] & r \leq 2^{\frac{1}{6}}\sigma 
			\\ 0 & r > 2^{\frac{1}{6}}\sigma \end{cases}.\] 

\begin{figure}
	\includegraphics[clip]{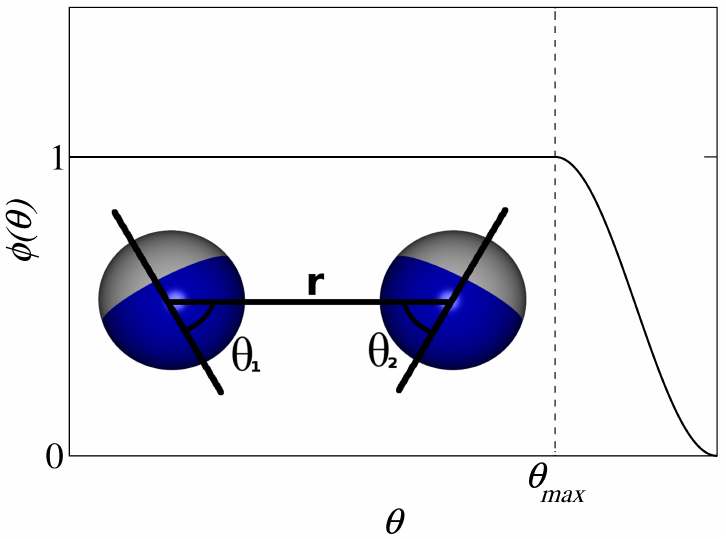}
	\caption{\label{fig:phi}(Color online) Sketch of the angular dependence of the inter-particle potential $\phi\left(\theta\right)$.  The dark side represents the hydrophobic region and the light side represents the hydrophilic.}
\end{figure}

$V_{\rm att}(r)$ accounts for the attraction between the hydrophobic \emph{surfaces} on the particles.
If $r_s = |r - \sigma |$ is the distance between the particle surfaces,
\begin{equation}
V_{\rm att}(r) = 4 \epsilon\Biggl[\left(\frac{\sigma/2}{r_s+\sigma/2\times2^{1/6}}\right)^{12}
\left.-2\left(\frac{\sigma/2}{r_s+\sigma/2\times2^{1/6}}\right)^6\right] \,,
\end{equation}
and it extends up to $r=1.5\sigma$. Finally,
$\phi\left(\theta\right)$ is a smooth step function that modulates the angular dependence of the potential, and 
is equal to 1 within the region $\theta \leq \theta_{\rm max}$ and decays  to 
zero following the expression $\cos^2(\pi(\theta-\theta_{\rm max})/(2\theta_{tail}))$ at the tail
of the angular range, i.e when $\theta_{\rm max}\leq\theta\leq\theta_{\rm max}+\theta_{tail}$ 
(with $\theta_{tail}=10^{\circ}$). This particular value of $\theta_{tail}$ has been selected to generate a sufficiently smooth potential at the Janus interface.
See Fig.~\ref{fig:phi} for an illustration of $\phi\left(\theta\right)$.

The parameters set in our model are compatible with colloidal particles of size $(200-400) nm$ in aqueous solution kept at 
a salt concentration sufficiently large to screen the electrostatic repulsion between the hydrophilic regions of any two particles~\cite{Hong2}.

The system is evolved using Molecular Dynamics simulations with a Langevin thermostat at constant room temperature, $T$, in a cubic box with periodic boundary conditions. Our system contains $N=10^3$ particles kept at a constant volume fraction $\phi=0.01$.  {We have chosen this concentration because it is comparable with those used in experimental studies on Janus Particles~\cite{Hong2}, so that our work could have a grounding experimental reference that our results could be compared to for $\theta_{\rm max}\simeq 90^{\circ}$.
Each simulation runs for  a minimum of $10^7$ steps with a time step $\delta t=0.001$. 
All quantities in this paper are expressed in standard dimensionless units.

\section{Results and discussion}
Our goal is to understand how particles condense into stable three-dimensional aggregates via the process of self-assembly, and how the specificity of the  geometry of the interaction is reflected in the final structure.
Figure 2 reports one of the main results of our simulations.  
It shows a diagram indicating the self-assembly lines separating the  structures obtained for different values of $\theta_{\rm max}$ and $\varepsilon$, with a typical resolution of one degree for $\theta_{\rm max}$, and 0.1$k_{\rm B}T$ for the binding energy.

\begin{figure}
   \subfigure{
       \includegraphics[width=3.28in,clip]{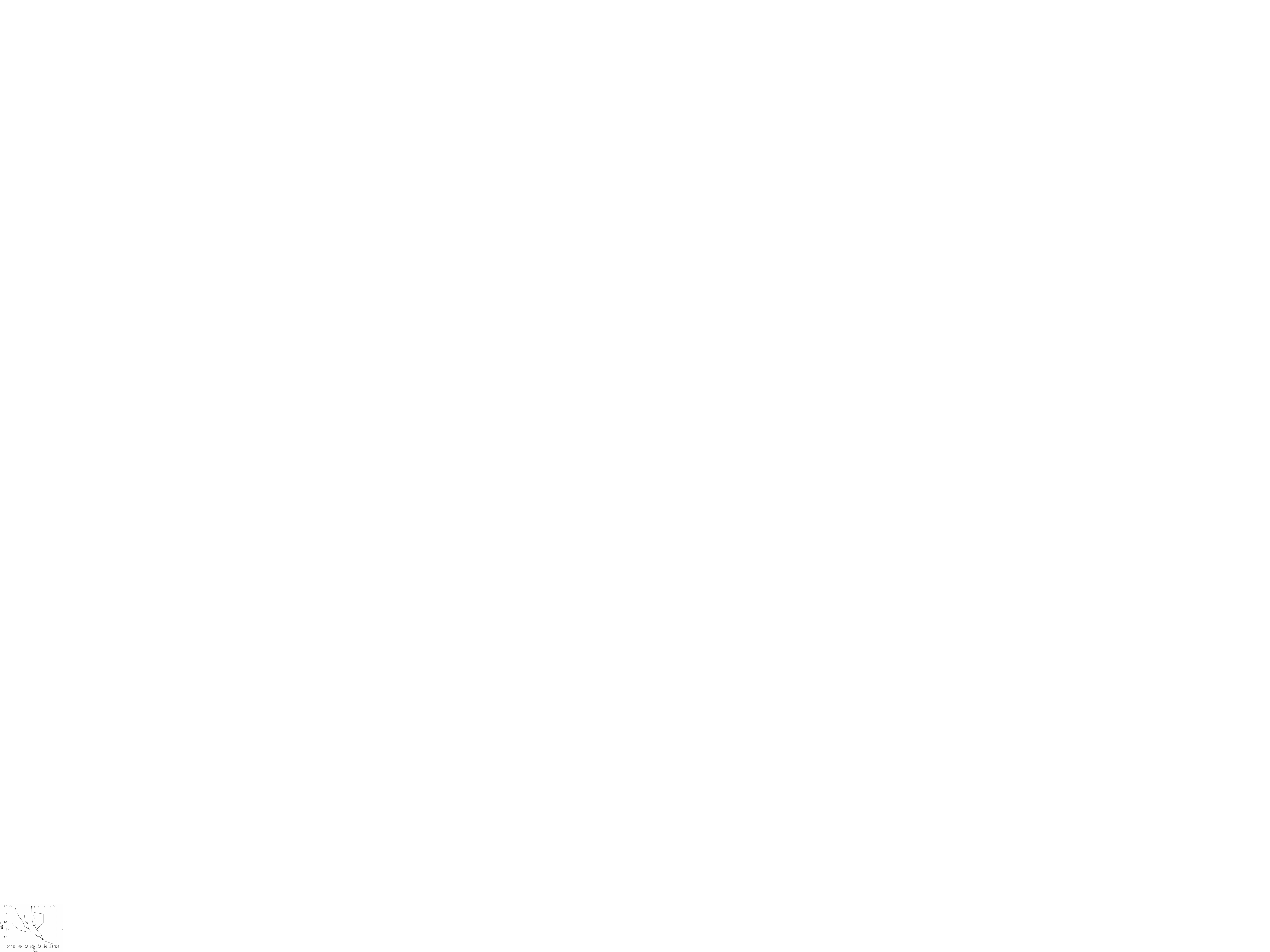}
       \put(-150,50){\bf{gas}}
       \put(-180,110){\bf{(a)}}
       \put(-170,150){\bf{(b)}}
       \put(-140,135){\bf{(c)}}
       \put(-105,160){\bf{(d)}}
       \put(-106,163){\vector(-1,0){10}}
       \put(-105,120){\bf{(e)}}
       \put(-60,120){\bf{(f)}}
       \put(-22,130){\bf{(g)}}
   }
   \subfigure{
       \includegraphics[width=1.64in,clip]{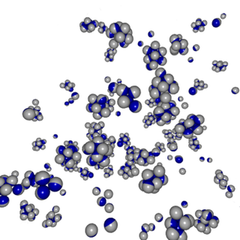}\label{fig:meso}
       \put(-115,87){\textcolor{black}{\bf{(a)}}}
   }
   \subfigure{
       \includegraphics[width=1.64in,clip]{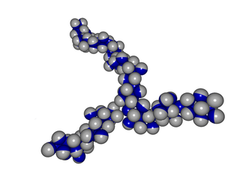}\label{fig:patch90}
       \put(-115,87){\textcolor{black}{\bf{(b)}}}
   }
   \subfigure{
       \includegraphics[width=1.64in,clip]{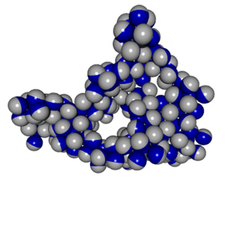}\label{fig:patch95}
       \put(-115,87){\textcolor{black}{\bf{(c)}}}
   }
   \subfigure{\includegraphics[width=1.64in,clip]{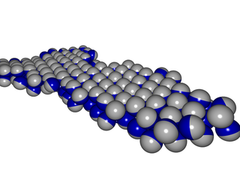}\label{fig:patch100}
       \put(-115,87){\textcolor{black}{\bf{(d)}}}
   }
   \subfigure{\includegraphics[width=1.64in,clip]{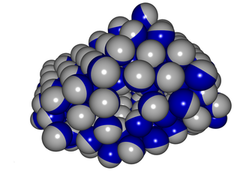}\label{fig:patch105}
       \put(-115,77){\textcolor{black}{\bf{(e)}}}
   }
   \subfigure{\includegraphics[width=1.64in,clip]{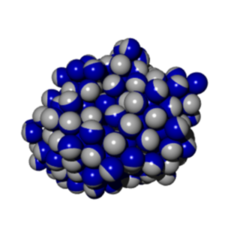}\label{fig:patch110}
       \put(-115,77){\textcolor{black}{\bf{(f)}}}
   }
   \subfigure{\includegraphics[width=1.64in,clip]{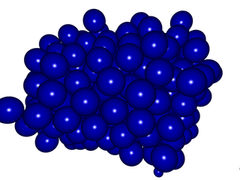}\label{fig:patch180}
       \put(-115,77){\textcolor{black}{\bf{(g)}}}
   }
   \caption{\label{fig:patchsize}(Color online) Self-assembly diagram of amphiphatic colloidal particles as a function of binding energy 
   $\varepsilon$ and size of  hydrophobic region  $\theta_{\rm max} $. Region (a) is populated by small micellar-like clusters containing
   4 to 13 particles. Region (b) contains branched  worm-like aggregates. In region (c) we find self-connected worm-like aggregates.
   Region (d) delimits flat hexagonal bilayers. In region (e) we find faceted hollow cages. Region (f) is populated by fluid amorphous 
   aggregates. Finally, in region (g) we find large clusters with FCC order.}
\end{figure} 

As expected, a rich variety of structures arises depending on the position of the 
dividing surface $\theta_{\rm max}$, and particles' binding energy, $\varepsilon$. Notice that,
consistent with recent numerical simulations on self-assembly of viral capsids~\cite{chandler},
and chaperonins~\cite{geissler},
 intermediate ordered  aggregates are extremely sensitive to the size of the hydrophobic region and self-assembly
occurs in a very narrow region.
At low binding energy the system is in a gas state.
For small values of $\theta_{\rm max}$ and moderate values of $\varepsilon$ (Fig.~\ref{fig:patchsize}(a)) particles aggregate into small  clusters of 4 to 13 particles (meso-particles)  {including icosohedral structures like those seen in \cite{wilber}}.
For $\theta_{\rm max} \sim 90^{\circ}$, self-assembly yields worm-like
extended structures (Fig.~\ref{fig:patchsize}(b)), as has been observed previously for Janus
 spheres~\cite{Hong2}. As the angular location of the hydrophobic region increases from $90^{\circ}$ to
$180^{\circ}$  we find, in order of appearance,  self-connected worm-like structures (Fig.~\ref{fig:patchsize}(c)), 
flat-crystalline bilayers  (Fig.~\ref{fig:patchsize}(d)), faceted hollow cages (Fig.~\ref{fig:patchsize}(e)) {(similar to structures seen in the self-assembly of cone-shaped particles~\cite{glotzer3,glotzer4}, yet not restricted to specific magic numbers of particles.)}, amorphous fluid  blobs (Fig.~\ref{fig:patchsize}(f)), and finally FCC crystals (Fig.~\ref{fig:patchsize}(g)).

Apart from the possible relevance of phases (d) and (e) for practical applications,
we want to point out that none of these structures arises following a  simple particle-by-particle
 growth mechanism, but rather by  two-to-three step hierarchical self-assembly.
The first step typically involves the formation of   {more- or less-structured} meso-particles (depending on $\theta_{\rm max}$). Next, small clusters organize into {either}  extended worm-like aggregates, which  {then coalesce} or deform to produce the final structure,  {or into larger fluid clusters that,  {once beyond some threshold size, spontaneously organize} into structured aggregates}.

It is of particular interest to discuss in some detail the dynamical pathway of structure formation relative
to phases (c), (d) and (e)  as they all form via a complex three step mechanism (the dynamical pathway leading to structures in region (b) is identical to what was found in reference ~\cite{Hong2} and we refer the reader to that paper for a thorough description.)
Surprisingly, the common precursor  to all of them is the worm-like structure stable at $\theta_{\rm max}\sim 90^{\circ}$.
Self-connected worm-like aggregates  are a consequence of the improved flexibility of  the 
worm-like structures. As $\theta_{\rm max}$ increases, so does the ability of particles 
to rotate about their axes. The net result is that the branching ends of the clusters begin to connect,
thus forming topologically nontrivial aggregates.

To quantify the statistical difference between the aggregates found in  regions (b) and (c),
we measure their average radius of gyration 
$R_{\rm G}=\left ( 1/N_c\sum_{i=1}^{N_c} <(\vec{r}-r_{\rm cm})^2> \right )^{1/2}$  as a function of cluster size. Results are shown in Fig.~\ref{gyration}, and clearly indicate how  
as $\theta_{\rm max}$  increases, the typical size of the aggregate decreases accordingly, resulting in more
compact structures. The low statistics for large clusters prevent us from making any meaningful 
estimate of the chains size exponent.
\begin{figure}
	\includegraphics{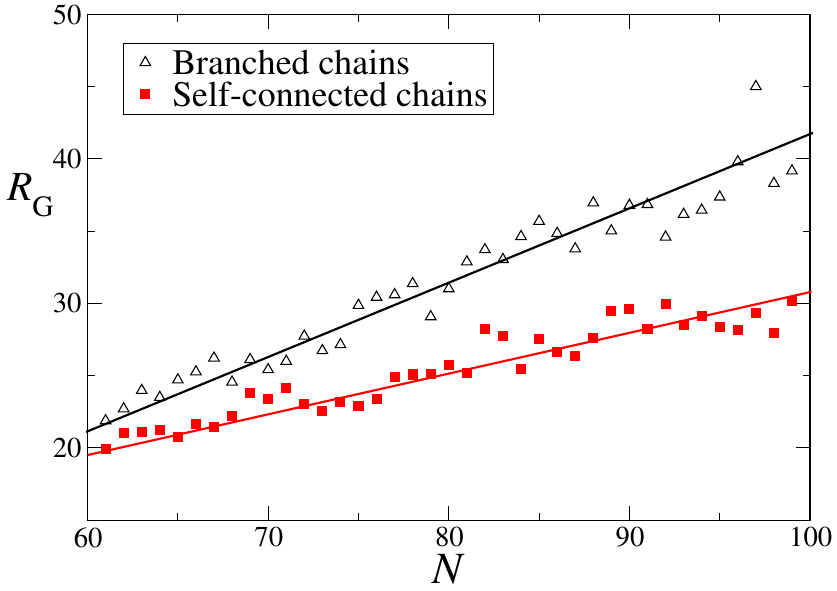}
		\caption{(Color online) Radius of gyration as a function of aggregate size $N$ for worm-like clusters in region (b) and (c) of Fig~\ref{fig:patchsize}. The lines are guides to the eye.}
		\label{gyration}

\end{figure}
We also measure 
the angular probability distribution function, $P(\cos(\alpha))$, between neighboring particles in  clusters containing at least 50 particles. Fig.~\ref{fig:p_of_alpha} shows  $P(\cos(\alpha))$ in regions (b), (c) and (d).
$P(\cos(\alpha))$ shows a clear double peaked shape in region (d), where the aggregates assume a planar bilayer configuration, but the more intriguing difference is between region (b) and (c), highlighted in the insert of the figure. Region (b) is characterized by a peak at $\cos(\alpha)\sim -1$, and less distinguishable peaks at   $\cos(\alpha)\sim \pm 0.74$ and  $\cos(\alpha)\sim -0.5$; overall there is a large probability for all possible orientations. These data suggest that  branched worm-like aggregates have a roughly circular cross-section with particles oriented in a disordered fashion, but for a slight preference for a few selected angles (remnants of the meso-particle structures they self-assembled from), and an anti-parallel neighbor opposite to most particles. In contrast, region (c) is characterized by a more distinct double-peaked function, with each peak close to $\pm 1$. This is consistent with a cross section that has flattened with respect to structures in region (b), and indicates that each branch of the aggregate is acquiring bilayer-like features. The dashed line separating regions (b) and (c)  shows where in the diagram the strings acquire sufficient flexibility to begin to form complex self-connected aggregates.   

This observation is quite interesting when related to the dynamic pathway leading to phase (d). In fact, bilayers are formed by either coalescence of co-planar loops, as shown in Fig.~\ref{fig:loops}, or by branch alignment. This transition  occurs at $\theta_{\rm max}\simeq 110^o$, and it is driven by  the large energy gain attained by the aggregates when each particle surrounds itself with the suddenly increased number of neighbors compatible with the enlarged hydrophobic region. This transition is quite sharp and represents a beautiful example of how a very small perturbation of the geometry of the local interaction can lead to completely different macroscopic structures.

Finally, formation of finite-size faceted capsids does not occur by self-assembly of  mis-oriented, disjoined bilayers, but via a mechanism which, once again,  involves multiple steps. As we increase the size of the particles' hydrophobic region, short worm-like clusters, which are now very flexible, immediately fold onto themselves to form small,  amorphous fluid blobs. We find that these blobs tend to remain fluid-like, whereas larger ones morph into faceted  cages via a mechanism similar to that described 
for the formation of bilayers. Fusion of fluid blobs, as illustrated in 
Fig.~\ref{fig:snapshots}, is the main mechanism through which large blobs, which eventually turn into  faceted cages, are generated.
The dashed line between regions (d) and (e) indicates the onset of cage formation; however, a non-negligible number of planar aggregates are found to coexist with the faceted cages
in region (e). It is worth stressing that these faceted cages can be considered as colloidal analogs of lipid vesicles; they are hollow,  their inner walls are hydrophilic, and one could in principle consider
using them as possible drug carriers, with edges and corners presenting convenient locations for outer surface tagging.

A statistical analysis of our data correlating  size and structure for non-planar aggregates in region (e) indicates a clear preference for large clusters to develop into  faceted hollow cages (see Fig.~\ref{fig:facetvsnon}). What sets the onset cluster size for this transformation is a complex compromise between the geometric constraints imposed by the interparticle potential; the energy gain to close-pack particles in a bilayer, which grows with the number of particles 
$N$; the energy cost for sides and corners, which have on average fewer neighbors than in a fluids state and have an energy cost which grows as $ N^{1/2}$ and $ N^{0}$ respectively; and finally the entropy loss due to particle ordering. Clearly, as $N$ increases, at sufficiently large binding energy,  planar configurations become the most stable, and this results  in surface faceting.

\begin{figure}
	\includegraphics{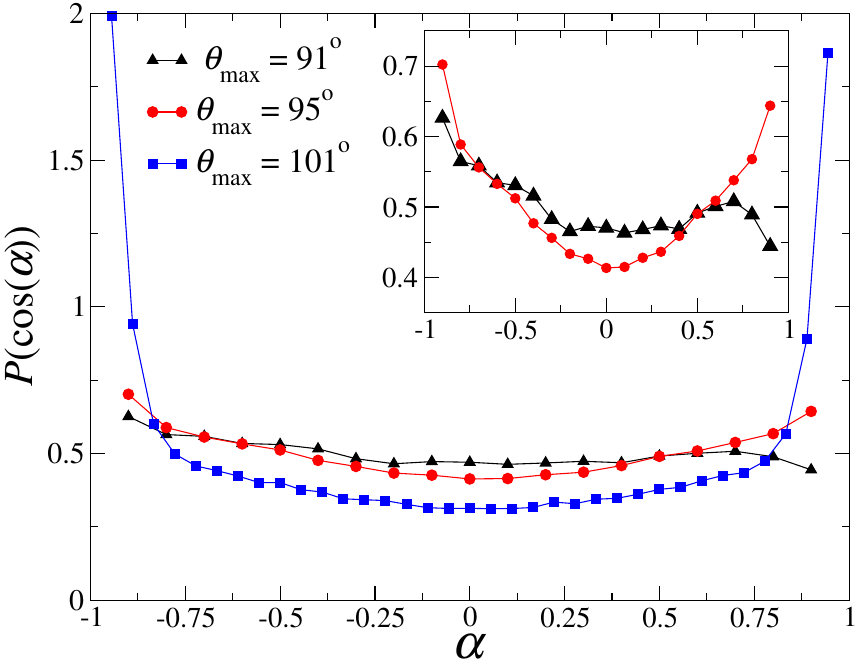}
	\caption{(Color online) Angular probability distribution function, $P(\cos(\alpha))$, between neighboring particles in  aggregates 	
	 containing at least 50 particles in region (b$\|\theta_{\rm max}=91^{\circ}$), 
	 (c$\|\theta_{\rm max}=95^{\circ}$), and (d$\|\theta_{\rm max}=101^{\circ}$), 
	 of the phase diagram. The insert highlights $P(cos(\alpha))$ 
	 in region (c) and (d). The data are averaged over 10 different clusters of each kind.}
	\label{fig:p_of_alpha}
\end{figure}
 \begin{figure}

\includegraphics[width=1.8in,clip]{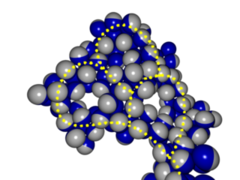}
\includegraphics[width=1.8in,clip]{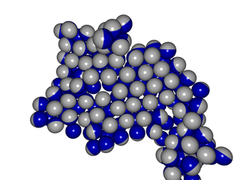}
\includegraphics[width=1.8in,clip]{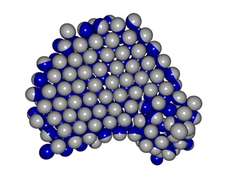}
		\caption{(Color online) Sequence of three snap-shots from our simulations showing the mechanism of  bilayer formation via coalescence of three coplanar loops.}
		\label{fig:loops}
\end{figure}

\begin{figure}
	\subfigure{
		\includegraphics[width=1.6in,clip]{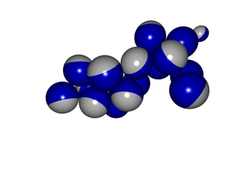}\label{fig:shota}
		\put(-115,77){\textcolor{black}{(1)}}
	}
	\subfigure{
		\includegraphics[width=1.6in,clip]{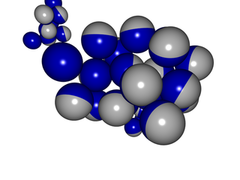}\label{fig:shotc}
		\put(-115,77){\textcolor{black}{(2)}}
	}
	\subfigure{
		\includegraphics[width=1.64in,clip]{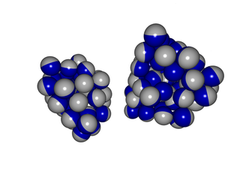}\label{fig:shot1}
		\put(-115,77){\textcolor{black}{(3)}}
	}

	\subfigure{
		\includegraphics[width=1.64in,clip]{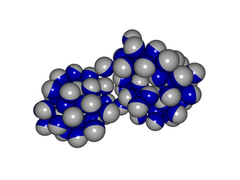}\label{fig:shot2}
		\put(-115,77){\textcolor{black}{(4)}}
	}	
	\subfigure{\includegraphics[width=1.64in,clip]{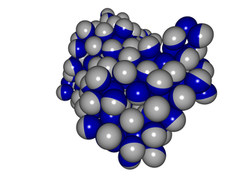}\label{fig:shot5}
		\put(-115,77){\textcolor{black}{(5)}}
	}
	\subfigure{\includegraphics[width=1.64in,clip]{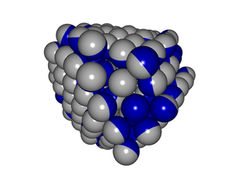}\label{fig:shot6}
		\put(-115,77){\textcolor{black}{(6)}}
	}
	\caption{(Color online) \label{fig:snapshots}Sequence of six snap-shots (1$\rightarrow$6) from our simulations showing faceted cage formation
	via fast folding of short worm-like clusters (1-2) and subsequent fusion of fluid blobs (3-6).}
\end{figure}

\begin{figure}
	\includegraphics{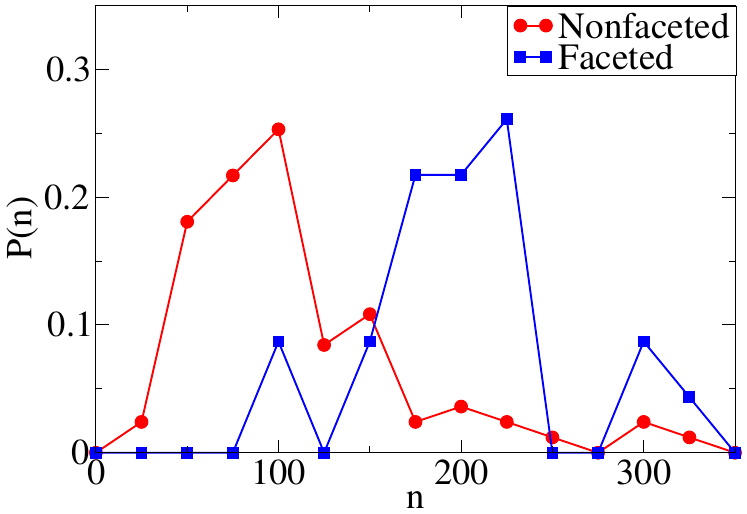}
	\caption{(Color online) Probability distribution function $P(n)$ as a function of cluster size $n$ for fluid and faceted aggregates
	in region (e) of the phase diagram.}\label{fig:facetvsnon}
\end{figure}

When $\theta_{\rm max}$ increases to even larger values, the geometric constraints imposed by
the inter-particle potential become less restrictive, and   planar bilayer configurations
become less stable. Phase  {(f)} is characterized by fluid, amorphous blobs which remain fluid at
all sizes. Large clusters, formed by the smooth fusion of smaller ones, tend to contain smaller sub-clusters in their interior. This is the first sign indicating that aggregates begin to acquire a three-dimensional character, which eventually leads to the formation of clusters with 
FCC  order, as inner and outer clusters begin to interact with each other.
The location of the fluid to FCC transition is at $\theta_{\rm max}\simeq 135^{\circ}$, and
was found by performing a structural analysis of the aggregates. Following~\cite{auer,ronchetti}, we identified particles whose local orientation is compatible to FCC ordering via 
a local bond order parameter based on spherical harmonics.
Given a particle  $i$, we consider
\begin{equation}
Q_{6m}(i)=  \frac{1}{N_b(i)}    \sum_{j=1}^{N_b(i)} Y_{6m}({\bf r}_{ij})\,,
\end{equation}
where $j$ runs over the $N_b(i)$ neighbors of  particle $i$, from which a rotationally invariant order parameter correlating the orientation 
of neighboring particles $i$ and $j$ can be defined as
\begin{equation}
{\bf q}_6(i)\cdot {\bf q}_6(j)= \sum_{m=-6}^6 Q_{6m}(i) \cdot Q^*_{6m}(j) 
\end{equation}
 Figure~\ref{fig:q6q6} shows the degree of crystallinity $\left<O\right>$ as 
a function of  $\theta_{\rm max}$ across the angular range $\theta_{\rm max}\in[120:180]$.
 $\left<O\right>$ is obtained by first averaging ${\bf q}_6(i)\cdot {\bf q}_6(j)$ over
 the neighbors of each particle $i$, and then by taking the average over  all crystalline particles
in a cluster. Clearly, once crystalline particles are formed, their degree
of order in a face-centered cubic crystal structure is independent of $\theta_{\rm max}$.
\begin{figure}
	\includegraphics{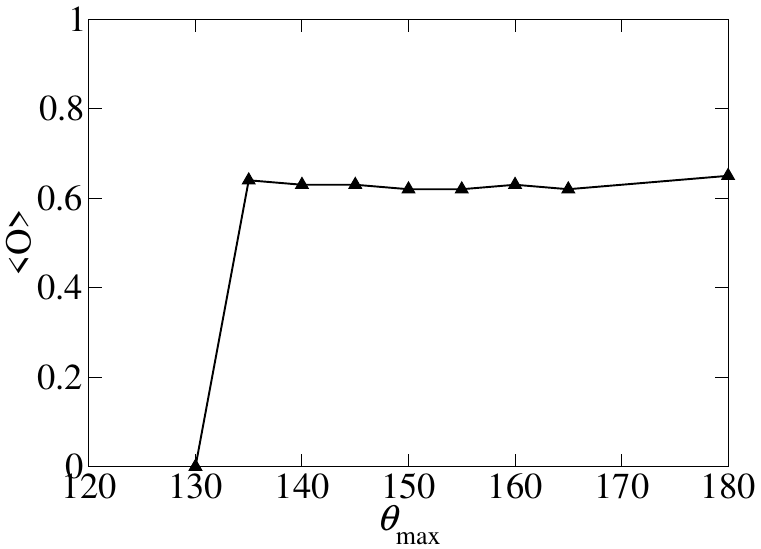}
	\caption{Degree of crystallinity of self-assembled aggregates as a function of hydrophobic area $\theta_{\rm max}$.}
	\label{fig:q6q6}
\end{figure}
The explanation of this behavior is purely geometrical. In fact, for sufficiently large $\theta_{\rm max}$,
the hydrophilic area on each particle becomes so small that can be positioned  among the 12 particles' contact
points resulting from an FCC crystal without affecting its structure.  Simple geometric considerations suggest that the
onset value should occur for an angular span of the hydrophobic region larger  {than} $150^{\circ}$. This value is compatible with
our result $\theta_{\rm max}\simeq 135^{\circ}$ because of the extra tail of $10^{\circ}$ in our definition of  the potential angular dependence.
A careful analysis of the model for decreasing values of $\theta_{\rm tail}$, not shown here, does indeed result
in a systematic shift of the crystallization onset $\theta_{\rm max}$ to larger values.
\begin{figure}
	\includegraphics[clip]{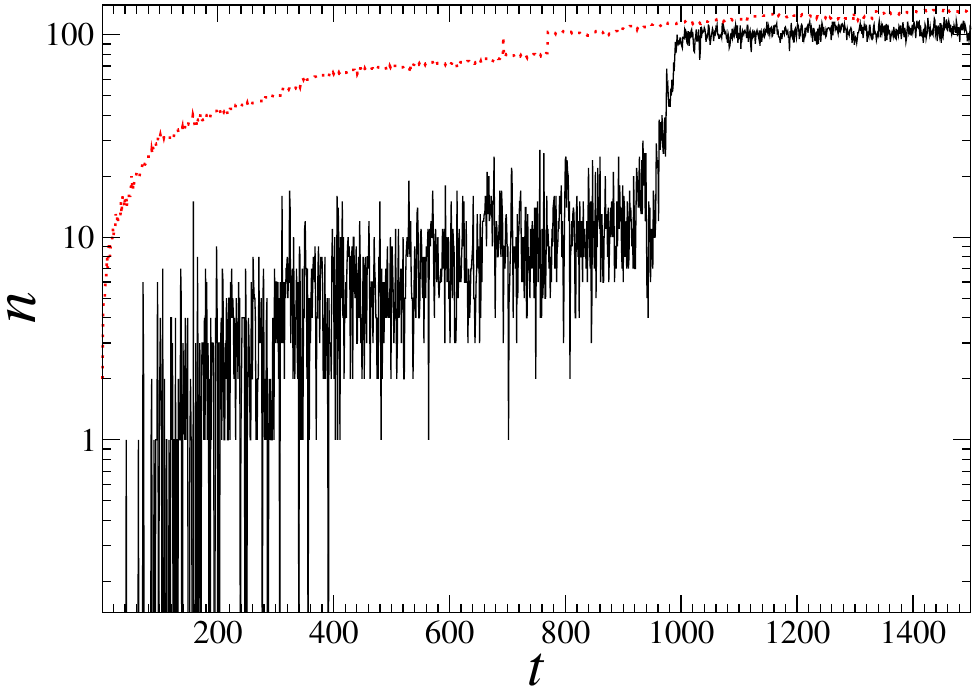}
	\caption{(Color online) \label{fig:nucleation} Linear-Log plot of the typical aggregate size $n$ 
as a function of time. The dotted line
shows the total number of particles in the aggregate, while the continuous line shows how many of those 
particles are tagged as crystal-like. These data were collected at $\theta_{\rm max}=160^{\circ}$.}
\end{figure}

At last, it is important to point out that, once again, the pathway leading to the formation 
of crystalline aggregates occurs in a two-step fashion. Particles first condense into a large, fluid
aggregate, and then crystallize from within the cluster via a standard nucleation process.
Figure~\ref{fig:nucleation} shows in the same plot the size dependence of the whole aggregate over time,
versus the number of crystalline particles in it. Clearly, crystal formation begins long after 
the aggregate is formed, and the large fluctuations in the initial stages of crystal growth show the typical 
signature of crystal nucleation. 
These results are compatible with nucleation studies of isotropic colloidal particles interacting via a short-range attractive potential first observed in~\cite{tenwolde}, and highlight the crucial role
played by meta-stable phases in the dynamics of crystal growth. In fact, we believe that this is the over-arching 
physics behind the rich dynamical phenomenology we find throughout this paper. Self-assembly proceeds, as predicted by Ostwald's step rule in the context of crystal nucleation~\cite{ostwald}, 
in a step-wise fashion that accounts for the  
complex free energy landscape containing  multiple meta-stable states.
Although we have not looked at the stability of the different phases found in our simulations,
planar and vesicular structures are also being  observed in the study of the equilibrium properties 
of a model system similar to ours~\cite{SciortinoX}. It is also worth mentioning that we expect the precise location of the phase boundaries to be somewhat sensitive to the particular choice of the angular potential. Unfortunately, this is very hard to characterize experimentally near the Janus interface and there is not a unique way of modeling that boundary. Our phase diagram is therefore intended to serve mostly as a guide for experimentalists.

\section{Conclusions}
In this paper we used molecular dynamics simulations to study the self-assembly pathway of spherical  amphiphatic colloidal particles.
We uncovered a wealth of different aggregates whose structures  span the three  dimensional spectrum. Specifically,
depending on the size ratio between the hydrophilic and  hydrophobic regions, particles self-assemble into small micellar clusters,
worm-like structures, planar bilayers, faceted and fluid cages, and finally FCC crystals.
We described   the hierarchical self-assembly pathway leading to most of these structures and discussed their connection to the geometry of local inter-particle
interactions. Finally, we made precise predictions  for the formation of hollow amphiphatic cages.

Although the morphology  of some of our aggregates can be predicted by simple geometric considerations, the dynamics leading to their formation 
is far less trivial, and may play a crucial role in the efficiency of the self-assembly process. We believe that, apart from trivial cases, any procedure 
attempting to design inter-particle interactions to target specific structures could greatly benefit from taking into account the dynamics of structure formation
in the design process.


\begin{thebibliography}{2}
\bibitem{Alberts} B. Alberts at al., \textit{Molecular Biology of the Cell}, 5th ed. (Garland Science, New York \& Oxford, 2008).
\bibitem{Watson} F.H.C. Crick , J.D. Watson, Nature {\bf 177}, 473 (1956).
\bibitem{Klug} A. Klug, D.L.D. Caspar, Advances in Virus Res {\bf{7}}, 225 (1960).
\bibitem{Sackmann} E. Sackmann, Canadian Journal of Physics {\bf 68},999, (1990).
\bibitem{Subramanian} G. Subramanian, V.N.Manoharan, J.D. Thorne, D.J. Pine,  Advanced Materials {\bf 11}, 1261 (1999).
\bibitem{DeVries} G.A. DeVries, et al., Science {\bf 315}, 358, (2007).
\bibitem{Schnablegger} M. Li, H. Schnablegger,S. Mann, Nature {\bf 402}, 393, (1999).
\bibitem{Hong} L. Hong, S. Jiang,S. Granick, Langmuir  {\bf 22}, 9495, (2006).
\bibitem{Weller} H. Weller, Philosophical Transactions of the Royal Society of London Series a-Mathematical Physical and Engineering Sciences  {\bf 361}, 229, (2003).
\bibitem{Hobbie} E. K. Hobbie et al., Langmuir {\bf 21}, 10284 (2005).
\bibitem{Israelachvili} J. N. Israelachvili, D. J. Mitchell, B. W. Ninham,  Journal of the Chemical Society-Faraday Transactions II {\bf 72}, 1525, (1976).
\bibitem{Zhang} T. Hu, R. Zhang,B. I. Shkovskii,  Physica a-Statistical Mechanics and Its Applications {\bf 387}, 3059, (2008).
\bibitem{Leckband} D. Leckband, J. Israelachvili, Quarterly Reviews of Biophysics {\bf 34}, 105 (2001).
\bibitem{Nagarajan} R. Nagarajan, E. Ruckenstein, Langmuir {\bf 7}, 2934 (1991).
\bibitem{glotzer2} S. C. Glotzer and M. J. Solomon, Nature Materials {\bf{6}}, 557 (2007).
\bibitem{Hong2} L. Hong, A. Cacciuto, E. Luijten and S. Granick, Langmuir {\bf 24}, 621 (2008).
\bibitem{glotzer} Z. Zhang and S. C. Glotzer, Nano Lett. {\bf{4}}, 1407 (2004).
\bibitem{sciortino} H. Liu et al., J. Chem. Phys. {\bf{130}}, 044902 (2009).
\bibitem{chandler} M. F. Hagan  and D. Chandler,  Biophys. J. {\bf 91}, 42 (2006).
\bibitem{geissler} S. Whitelam and P.L. Geissler, J. Chem. Phys. {\bf{127}}, 154101 (2007).
\bibitem{wilber} A. X. Wilber et al., J. Chem. Phys. {\bf{127}}, 085106 (2007).
\bibitem{glotzer3} T. Chen, Z. Zhang and S. C. Glotzer, PNAS {\bf{104}}, 717 (2007).
\bibitem{glotzer4} T. Chen, Z. Zhang and S. C. Glotzer, Langmuir {\bf{23}}, 6598 (2007).
\bibitem{auer} S. Auer and D. Frenkel, Annu. Rev. Phys. Chem. {\bf 55}, 333, (2004).
\bibitem{ronchetti} P. L. Steinhardt, D. R. Nelson, M. Ronchetti, Phys. Rev. B {\bf 28}, 784 (1983).
\bibitem{tenwolde} P. R. ten Wolde and D. Frenkel, Science {\bf 277}, 1975 (1997). 
\bibitem{ostwald} W. Z. Ostwald, Phys. Chem. {\bf 22}, 289, (1879).
\bibitem{SciortinoX} F. Sciortino, private communication. 
\end{thebibliography}
\end{document}